\documentclass[12pt]{article}
\usepackage{graphicx}
\usepackage{natbib}
\usepackage{astro}
\usepackage{amsmath,amsfonts,amssymb}
\long\def\symbolfootnote[#1]#2{\begingroup
\def\thefootnote{\fnsymbol{footnote}}\footnote[#1]{#2}\endgroup} 
\long\def\symbolfootnotemark[#1]{\begingroup
\def\thefootnote{\fnsymbol{footnote}}\footnotemark[#1]\endgroup} 
\long\def\symbolfootnotetext[#1]#2{\begingroup
\def\thefootnote{\fnsymbol{footnote}}\footnotetext[#1]{#2}\endgroup} 
\def\Msun{\rm{M_\odot}}
\def\Jupiter{J}
\def\earth{\oplus}
\def\microas{\mu{\rm as}}
\def\AU{AU}

\def\url#1{\texttt{#1}}

\begin{document}
\thispagestyle{empty}

%
%

{\center 
{\Large Observational Techniques for Detecting Planets in Binary Systems}\\
{Matthew W.~Muterspaugh, Maciej Konacki, Benjamin F.~Lane, and Eric Pfahl}
}


\section{Why Focus Planet Searches on Binary Stars?}

Searches for planets in close binary systems explore the degree to which 
stellar multiplicity inhibits or promotes planet formation.  
There is a degeneracy between planet formation 
models when only systems with single stars are studied---several mechanisms 
appear to be able to produce such a final result.  This degeneracy is lifted 
by searching for planets in binary systems; the resulting detections (or 
evidence of non-existence) of planets in binaries isolates which models may 
contribute to how planets form in nature.  Studying relatively close pairs of 
stars, where dynamic perturbations are the strongest, provides the most 
restrictive constraints of this type 
\citep[see, for example,][]{thebault2004, Pfahl2005, PfahlMute2006}.

In this chapter, we consider observational efforts to detect planetary 
companions to binary stars in two types of hierarchical planet-binary 
configurations:  first ``S-type'' planets which orbit just one of the 
stars, with the binary period being much longer than the planet's; 
second, ``P-type'' or circumbinary planets, where the planet 
simultaneously orbits both stars, and the planetary orbital period is 
much longer than that of the binary \citep{Dvorak1982}.  Simulations show 
each of these configurations has a large range of stable configurations 
\citep[see, e.g.,][also, this book, chapters ZZZZ]
{Benest2003, PL2003, PL2002, Broucke2001, 
holman1999, Benest1996, Benest1993, Benest1989, Benest1988, Rabl1988}.  


\section{S-Type Planets}

S-Type planets orbit just one of the stars in a binary, and the binary 
separation is much larger than that between the star and planet.  Some of the 
binaries are so widely separated (projected semimajor axis 
$a_b \gtrsim 1$ arcsecond) that they can be 
spatially resolved by ground-based telescopes without active image correction; 
for these, traditional planet-finding techniques can be used.  In fact, 
astrometric methods often perform best in this regime, as the secondary star 
serves as a convenient reference for the primary, and vice versa.  Here, 
astrometric and radial velocity (RV) programs are considered as the most 
versatile search methods.  (While transit searches might also be possible, 
these typically have very limited spatial resolutions, and the second star can 
act as a photometric ``contaminant.'')  When the binaries are not spatially 
resolved with simple imaging, modifications must be made to meet the 
measurement precisions required for detecting extrasolar planets.


\subsection{Wide Binaries}

From an observational standpoint, ``wide'' binaries are considered to be those 
that can be resolved by traditional (uncorrected) imaging techniques.  Due to 
atmospheric seeing, this sets the projected sky separation at larger than 
roughly one arcsecond.


\subsubsection{Dualstar Astrometry}

Interferometric narrow-angle astrometry \citep{shao92,col94}
promises astrometric performance at the 10-100 micro-arcsecond
level for pairs of stars separated by 1-60 arcseconds.  
The lower limit of the allowable binary separation for this technique is that 
the binary is resolved by the individual telescopes in the interferometer; 
the upper limit is set by the scale over which the effects of atmospheric 
turbulence are correlated.  This technique 
was first demonstrated with the Mark III interferometer for short 
integrations \citep{col94}, was extended to longer integrations and 
shown to work at the 100 micro-arcsecond
level at the Palomar Testbed Interferometer \citep[PTI, ][]{l00}.

However, achieving
such performance requires simultaneous measurement of the interferometric 
fringe positions of both stars, greatly complicating the instrument (two beam
combiners and metrology throughout the entire array are required). 
In addition, the instrumental baseline vector $\overrightarrow{B}$ connecting 
the unit telescopes must be
known to high precision ($\approx 100$ microns).

\begin{figure}[tbp]
\begin{center}
\includegraphics[height=3.5in]{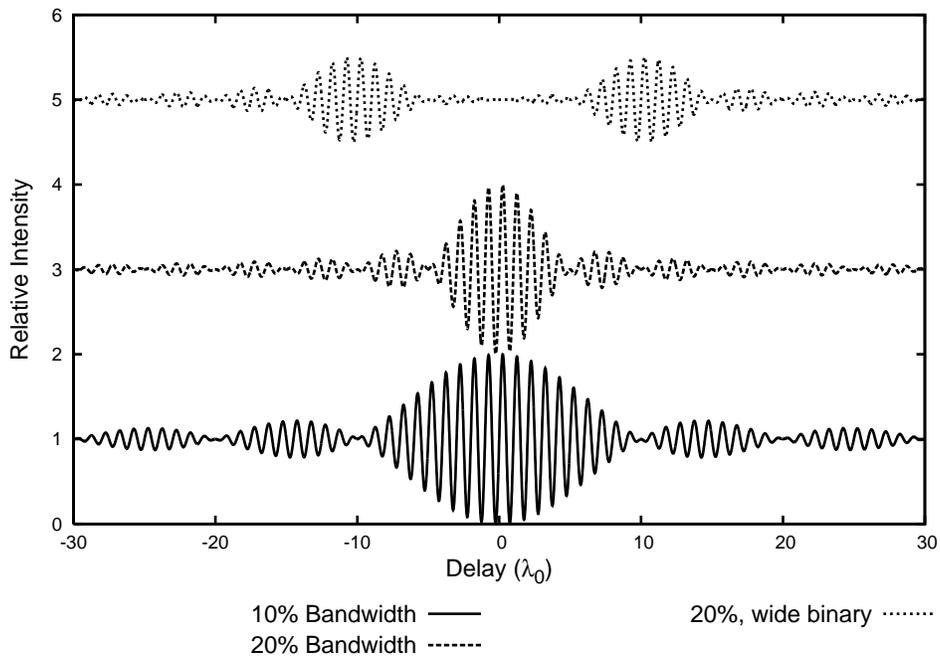}
\end{center}
\caption[Response of an Interferometer]
{\label{fig:fringes} 
The response of an interferometer.  The top two curves have 
been offset by 2 and 4 for clarity.  
The widths of the fringe packets are determined by the 
bandpass of the instrument, and the wavelength 
of fringes by an averaged wavelength of starlight.  
The top curve shows the intensity pattern 
obtained by observing two stars separated by a small 
angle on the sky---the observable is the distance between the 
fringe packets.}
\end{figure}

In an optical interferometer light is collected at two or more
apertures and brought to a central location where the beams are
combined and a fringe pattern produced on a detector.
For a broadband source of central wavelength $\lambda$ and
optical bandwidth $\Delta\lambda$ the fringe pattern is limited in extent and
appears only when the optical paths through the arms of the
interferometer are equalized to within a coherence length ($\Lambda =
\lambda^2/\Delta\lambda$). For a two-aperture interferometer,
neglecting chromatic dispersion by unequal air paths, 
the intensity measured at one of the combined
beams is given by
\begin{equation}\label{double_fringe}
I(x) = I_0 \left [ 1 + V \frac{\sin\left(\pi x/ \Lambda\right)}
{\pi x/ \Lambda} \sin \left(2\pi x/\lambda \right ) \right ]
\end{equation}
\noindent where $V$ is the fringe contrast or ``visibility'', which
can be related to the morphology of the source,
and $x$ is the optical path difference between arms of the
interferometer; see Fig.~\ref{fig:fringes}.  More detailed analysis
of the operation of optical interferometers can be found in {\it Principles of
Long Baseline Stellar Interferometry} \citep{Lawson2000}.

The location of the resulting interference fringes are
related to the position of the target star and the observing geometry
via
\begin{equation}\label{delayEquation}
d = \overrightarrow{B} \cdot \overrightarrow{S} + 
    \delta_a\left(\overrightarrow{S}, t\right) + c
\end{equation}
\noindent where $d$ is the optical path-length one must introduce between the 
two arms of the interferometer to find fringes (often called the ``delay''),  
$\overrightarrow{S}$ is the unit vector
in the source direction, and $c$ is a constant additional scalar delay
introduced by the instrument.  The term 
$\delta_a\left(\overrightarrow{S}, t\right)$
is related to the differential amount of path introduced by the atmosphere
over each telescope due to variations in refractive index.  

If the other quantities are known or small, measurement of 
the instrumental path 
length $d$ required to observe fringes determines the position of the star 
$\overrightarrow{S}$.
For a 100-m baseline interferometer, an astrometric precision of 10 $\mu$as
corresponds to knowing $d$ to 5 nm, a difficult but not impossible
proposition for all terms except that related to 
the atmospheric delay. Atmospheric turbulence, which changes over
distances of tens of centimeters and on millisecond timescales, forces
one to use very short exposures to maintain fringe contrast, and
hence limits the sensitivity of the instrument. It also severely limits
the astrometric accuracy of a simple interferometer, at least over
large sky-angles.

However, in narrow-angle astrometry one is concerned with a close pair
of stars, and the observable is a differential astrometric measurement, 
i.e.~one is interested in knowing the angle between the two stars 
($\overrightarrow{\Delta_s} = \overrightarrow{s_2} - \overrightarrow{s_1} $). 
The atmospheric turbulence is correlated over
small angles.  If the measurements of the two stars are simultaneous, or 
nearly so, the atmospheric term subtracts out making possible 
high precision ``narrow-angle'' astrometry. 

\begin{figure}
\begin{center}
\includegraphics[height=8.0cm]{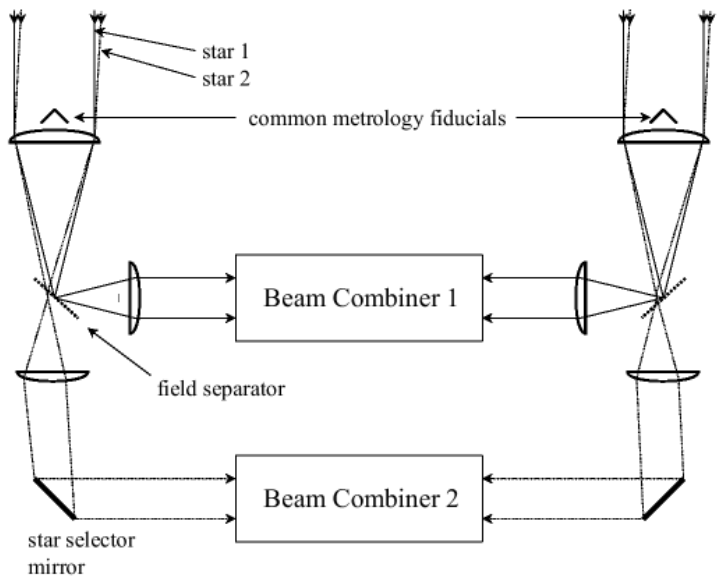}
\end{center}
\caption[Dual-Star Interferometer]
{\label{fig:dsm}
Schematic of splitting the light in a dualstar interferometer.
}
\end{figure}

The requirement that the target and reference stars be observed 
simultaneously results in a significant instrumental complexity, i.e. 
essentially two complete interferometers are required to share the 
same set of apertures (see Fig.~\ref{fig:dsm}).  
The splitting of light from the stars into 
two separate sets of delay lines, beam transport systems and beam combiners
is done in a ``dual-star module'' located just after the apertures,
with the split generally being accomplished using a beam-splitter.  
Considerable care must be taken in designing the system in order 
to avoid small pathlength measurement errors.

\begin{figure}
\begin{center}
\includegraphics[scale = 0.5]{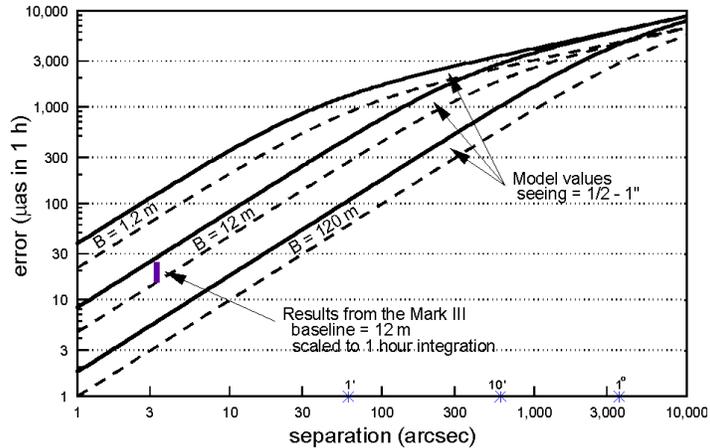}
\caption[Narrow-Angle Astrometric Precision]{\label{fig:naa}
Astrometric accuracy vs. star separation in a one-hour integration for
different baseline lengths.  Model atmospheres providing 1/2- and
1.0-arcsecond seeing are shown.  These results assume an infinite outer
scale, and better results are achieved when the baseline exceeds the
outer scale, as would be expected with a 100 m baseline at most sites.
Measurements with the Mark~III interferometer of a 3.3~arcsecond binary
star are consistent with the model. This figure is from \cite{shao92}.}
\end{center}
\end{figure}

The exact level of astrometric precision that can be achieved depends on 
many factors, including the separation of the target/reference pair, the 
size of the interferometric baseline and the levels and distribution of 
atmospheric turbulence. For a typical Mauna Kea seeing profile the astrometric 
precision is
\begin{equation}
\sigma_a \simeq 300\frac{\theta}{\sqrt{t}B^{2/3}}~{\rm arcsec}
\end{equation}
\noindent where $B$ is the baseline length in meters, $\theta$
is the target/reference separation in radians, and t is the integration 
time in seconds.  
For typical baselines of $\sim 100$ m, and an angular
separation of $\sim 30$ arcsecond implies an astrometric precision of 30
$\microas$ in an hour (see Fig.~\ref{fig:naa}).  

The magnitude of the astrometric signal of the star's motion about the 
center of mass (CM) between it and its planet is given by:
\begin{equation}\label{ast_reflex}
\Delta a_{CM} = 2 \frac{M_p}{M_s}a_p = 
\frac{M_p/M_\Jupiter}{M_s/\Msun}\frac{a_p}{524}.
\end{equation}
where $M_p$, $M_b$, $M_\Jupiter$, $\Msun$ are, respectively, masses of the 
planet,star, Jupiter, and Sun, and $a_p$ is the semimajor axis of the planet's 
orbit.

Thus, the minimum mass that can be detected is roughly
\begin{eqnarray}\label{ast_ref2}
M_p/M_\Jupiter & \gtrsim & 524 \frac{\sigma_a}{a_p} \frac{M_s}{\Msun}\\
              & \gtrsim & 0.1 \frac{\sigma_a/20\microas}{a_p/1\AU}
                          \frac{d}{10 {\rm pc}}\frac{M_s}{\Msun}\\
M_p/M_\earth   & \gtrsim & 1.6 \frac{\sigma_a/1\microas}{a_p/1\AU}
                          \frac{d}{10 {\rm pc}}\frac{M_s}{\Msun}
\end{eqnarray}
where $M_\earth$ is the mass of the Earth, and here $d$ is the distance 
to the target star.


\subsubsection{Radial Velocities}\label{wideRV}

When the stars in a binary can be spatially resolved without active image 
correction on ground based telescopes, the spectrum of each star can be 
recorded separately without contamination from the other, and the standard 
precision RV method described below can be used \citep[see, for example, ][]
{Campbell1988, Butler1996}.  Similarly, if the secondary is 
much fainter than the primary, precision RV might be performed on the brighter 
star as though it were single, though there is 
concern about the influence of the fainter lines.  
Several ($\sim 30$)exoplanet candidates in binaries 
have been discovered in this manner.  In some 
cases the stars were not previously known to be binaries, and their natures 
were only discovered by long-term RV trends or follow-up adaptive optics 
imaging.  Some of these efforts to detect planets in binary stellar 
systems include that of \cite{Toyota2005} for single-lined and wide binaries, 
that of \cite{Desidera2006b} targeting wide binaries, 
and the program targeting single-lined and wide binaries of \cite{Udry2004}.

The highest precision RV observations are obtained either from the $I_2$ 
(molecular iodine) absorption cell or the use of carefully designed 
spectrographs with fiber scrambling. In order to achieve a RV precision of 
$\sim 1 m\,s^{-1}$ 
an iodine absorption cell is used to superimpose a reference spectrum on 
the stellar spectrum (by sending a starlight through the cell). The  
spectrum provides a fiducial wavelength scale against which radial 
velocity shifts are measured. 

Thanks to its conceptual simplicity, the iodine technique is the most 
commonly adopted way to obtain precision radial velocities. Iodine 
absorption cells are available on many spectrographs---HIRES at 
the 10m Keck I (Keck Observatory), 
Hamilton at the 3m Shane (Lick Observatory), 
SARG at the 3.6m TNG (Canary Islands), UCLES at the 3.9m Anglo-Australian 
Telescope (Anglo-Australian Observatory), HRS at the 9m HET 
(McDonald Observatory), MIKE at the 6.5m Magellan 
(Las Campanas Observatory), UVES at the 8m Kueyen (Cerro Paranal), 
HDS at the 8.2m Subaru (National Astronomical Observatory of Japan) 
and many other---and are used for planet detections.

In the iodine absorption cell technique, the Doppler shift of a star spectrum
is determined by solving the following equation \citep{Marcy1992}
\begin{equation}
\label{i2::}
I_{obs}(\lambda) =
[I_{s}(\lambda+\Delta\lambda_{s})\,T_{I_{2}}(\lambda+\Delta\lambda_{I_{2}})]
\,\otimes\,PSF
\end{equation}
where $\Delta\lambda_{s}$ is the shift of the star spectrum,
$\Delta\lambda_{I_{2}}$ is the shift of the iodine transmission
function $T_{I_{2}}$, $\otimes$ represents a convolution, and
$PSF$ a spectrograph's point-spread function.  
The parameters $\Delta\lambda_{s},
\Delta\lambda_{I_{2}}$ as well as parameters describing the PSF
are determined by performing a least-squares fit to the observed
spectrum, $I_{obs}$, as seen through the iodine cell.  To this
end, one also needs a high SNR star spectrum taken without
the cell, $I_{s}$, which serves as a template for all the
spectra observed through the cell, as well as the I$_2$ transmission
function, $T_{I_{2}}$, obtained with the Fourier 
Transform Spectrometer at the Kitt Peak National Observatory.
The Doppler shift of a star spectrum is then given by
$\Delta\lambda = \Delta\lambda_{s} - \Delta\lambda_{I_{2}}$.

The velocity reflex amplitude of a star due to an unseen companion is given by 
\begin{eqnarray}\label{vel_reflex}
\Delta v_b & = & 2 \frac{2 \pi a_p \sin i_p}{P_p}\frac{M_p}{M_s+M_p}
            = \frac{2 \sqrt{G} M_p \sin i_p }{\sqrt{\left( M_s+M_p \right) a_p}}\nonumber\\
           & = & 56.9\,{\rm m\,s^{-1}} \times \frac{ \left( M_p \sin i_p /M_\Jupiter \right)}
                 {\sqrt{\left( \left( M_s+M_p \right)/\Msun\right)\left(a_p/1\rm{AU} \right)}}
\end{eqnarray}
\noindent where $P_p$ is the period of the planet's orbit, 
$G$ is the gravitational constant, and 
$i_p$ is the inclination of the planet's orbit.

For an RV precision of $\sigma_{rv}$, the 
minimum mass that can be detected is roughly
\begin{eqnarray}
M_p \sin i_p / M_\Jupiter & \gtrsim & 0.018 
\frac{\sigma_{rv}}{1{\rm m\,s^{-1}}} 
{\sqrt{\left( \left( M_s+M_p \right)/\Msun\right)\left(a_p/1\rm{AU} \right)}}\\
M_p \sin i_p / M_\earth & \gtrsim & 5.6 
\frac{\sigma_{rv}}{1{\rm m\,s^{-1}}} 
{\sqrt{\left( \left( M_s+M_p \right)/\Msun\right)\left(a_p/1\rm{AU} \right)}}.
\end{eqnarray}


\subsubsection{Observational Precisions}

\begin{figure}[tbp]
\begin{center}
\includegraphics[height=3.5in]{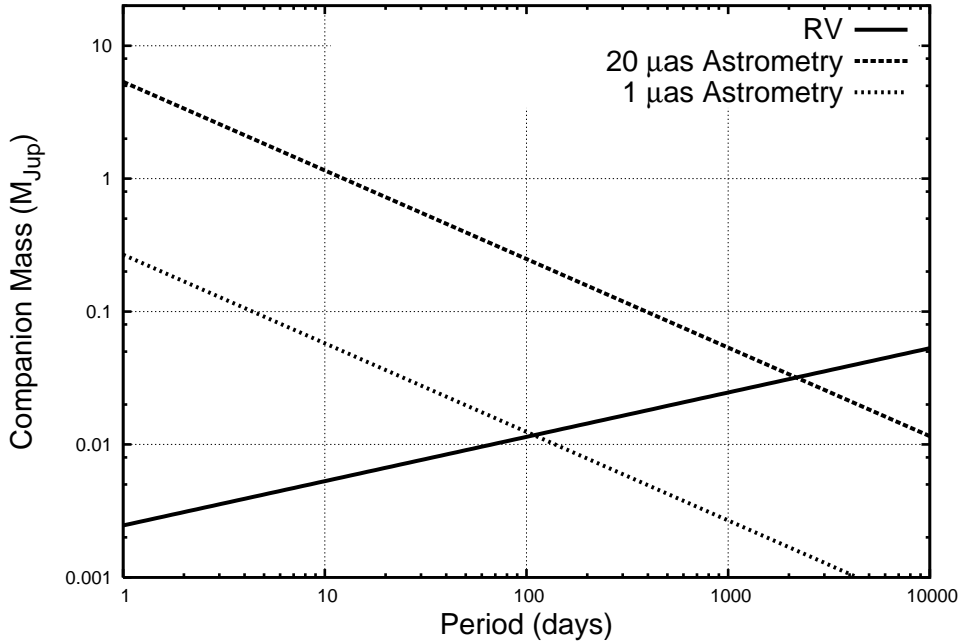}
\end{center}
\caption[Sensitivity to S-Type Planets in Wide Binaries]
{\label{fig:wideS} 
Sensitivity to S-Type planets in wide binaries, comparing 
astrometric and radial velocity techniques.  All calculations 
assume solar mass stars; astrometric sensitivity assumes a distance 
of 10 pc to the target system.
}
\end{figure}

Astrometry is most sensitive to long period planets, RV to short period ones.  
Figure \ref{fig:wideS} shows the companion masses 
one can detect for each method, assuming 
20 $\microas$ ground-based astrometry, 1 $\microas$ space-based astrometry, 
and $1 {\rm m\, s^{-1}}$ RV precisions.

\subsection{Close Binaries}

Radial velocity surveys for extrasolar planets have been restricted
largely to stars within $\simeq$100\,pc, and the majority of detected
exoplanets are at distances of 10--50\,pc.  Therefore, our
observational definition of a ``wide'' binary corresponds to projected
orbital separations of $\gtrsim$50\,AU.  More compact systems form the
complementary class of ``close'' binaries.  From a theoretical
standpoint, binaries with semimajor axes of $\lesssim$50\,AU pose important
challenges to standard ideas about the formation of giant planets. 

Imagine a protoplanetary disk around one star in a newly formed
binary.  Suppose that the orbit has semimajor axis $a$ and
eccentricity $e$, and assume, for simplicity, that the stars have
equal masses.  The tidal gravitational field of the companion star
truncates the disk at a radius of $R_t \simeq 0.26 a(1-e^2)^{1.2}$ 
\citep[e.g.~][]{Pichardo2005}.  If $R_t\lesssim 3\,{\rm AU}$ 
(i.e., inside the so-called ``ice line''), it seems
unlikely that icy grains could form and grow into planetesimals, thus
precluding the embyonic stage of giant planet formation in the
core-accretion scenario \citep[e.g.,][]{Liss1993}.  In more
extreme cases, when $R_t \lesssim 1\,{\rm AU}$, there may be insufficient
material in the disk to yield a Jovian-mass planet
\citep{Jang-Condell2006}.  Even when $R_t$ is as large as
$\simeq$10\,AU, stirring of the disk by the tidal field and the
thermal dissipation of spiral waves may inhibit planetesimal
formation, as well as stabilize the disk against fragmentation
\citep[][]{Nelson2000, thebault2004, Thebault2006}.  In this case, it
may be that neither the core-accretion picture nor gravitational
instability \citep[e.g.,][]{Boss2000} are accessible modes of giant
planet formation.  We adopt $R_t = 10\,{\rm AU}$ ($a \lesssim 40\,{\rm AU}$
for modest $e$) as a fiducial upper limit for which Jovian planet
formation is significantly perturbed and perhaps strongly inhibited.
This provides a simple theoretical definition of a close binary that
roughly matches our observational measure.

A handful of planets in binaries with $a \lesssim 20\,{\rm AU}$ have
already been discovered (see Table \ref{tableclose}).  
The tightest of these systems
\citep[HD 188753][]{Konacki2005} has a periastron separation of only
$\simeq$6\,AU, which seems severely at odds with the conventional lore
on Jovian planet formation.  \citet{Pfahl2005} and
\citet{Portegies_Zwart_2005} suggested that the planetary host star may
have been acquired in a dynamical exchnage interaction in a star
cluster {\em after} the planet formed, thus circumventing the
complicating factors listed above.  As most stars are born in
clustered environments, one wonders how often dynamics can account for
planets in close binaries.  This idea was explored in
\citet{PfahlMute2006}, where it was found that exchange
interactions can account for only $\sim$0.1\% of close
binaries hosting planets.  However, the (admittedly small) sample of systems in
Table~1 seems to indicate that a larger fraction of $\sim$1\% of close
binaries harbor giant planets (see Pfahl \& Muterspaugh 2006 for
details), and perhaps planets do somehow form frequently in these
hostile environments.  It crucial that we begin to develop a census of
planets in close binaries in order to test the different theories
about planet formation and dynamics.

\begin{table}[tbp]
\begin{center}
Close Binaries with Planets
\begin{tabular}{cccccc}
\hline
\hline
Object            & $a_b$ (AU) & $e$       & $M_1/M_2$  & $R_t$ (AU) & Refs \\
\hline
HD 188753         & 12.3       & 0.50      & 1.06/1.63  & 1.3        & 1 \\
$\gamma$ Cephei   & 18.5       & 0.36      & 1.59/0.34  & 3.6        & 2, 3 \\
GJ 86             & $\sim$20   &           & 0.7/1.0    & $\sim$5    & 4, 5, 6 \\
HD 41004          & $\sim$20   &           & 0.7/0.4    & $\sim$6    & 7 \\
HD 196885         & $\sim$25   &           & 1.3/0.6    & $\sim$7    & 8 \\ 
\hline
\end{tabular}
\end{center}
\caption[Close Binaries with Planets]
{ \label{tableclose}
When no eccentricity is given, only the projected binary separation is known.  
$M_1/M_2$ is the planetary host mass divided by companion mass.  
In HD 188753, the secondary is a binary with semimajor axis 0.67 AU.  
In GJ 86, the secondary is a white dwarf; to estimate the tidal truncation 
radius $R_t$, an original companion mass of $1\Msun$ is assumed.  
(1) \cite{Konacki2005};
(2) \cite{Campbell1988}
(3) \cite{hatzes2003};
(4) \cite{Queloz2000} 
(5) \cite{Mugrauer2005} 
(6) \cite{Lagrange2006}
(7) \cite{Zuc2004} 
(8) \cite{Chauvin2006b}
}
\end{table}


\subsubsection{PHASES Astrometry}

The dualstar astrometry method can be modified for use when the binaries are 
so close that the individual telescopes of an interferometer cannot resolve 
the pair \citep{LaneMute2004a}.  
The interferometer itself overresolves the binary, as in 
Fig.~\ref{fig:fringes}; its high spatial resolution then allows for precision 
astrometric measurements.

In this mode, the small separation of the
binary results in both components being in the field of view of
a single interferometric beam combiner. The fringe positions are
measured by modulating the instrumental delay with an amplitude large
enough to record both fringe packets.  

\begin{figure}[]
   \centerline{\includegraphics[height=2.5in]{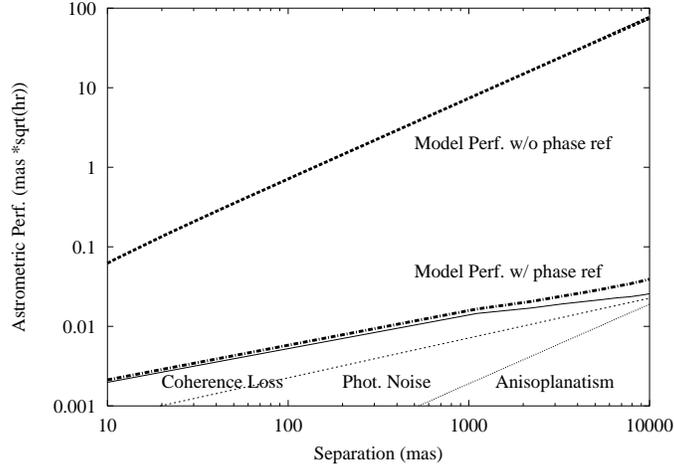}}
\caption{\label{fig:expected} The expected narrow-angle astrometric
performance in mas for the phase-referenced fringe-scanning
approach, for a fixed delay sweep rate, and an interferometric
baseline of 110 m.  
Also shown is the magnitude of the temporal loss of coherence effect
in the absence of phase referencing, illustrating why stabilizing the
fringe via phase referencing is necessary.  }
\end{figure}

However, since the fringe position measurement of the two stars is no
longer truly simultaneous it is possible for the atmosphere to
introduce path-length changes (and hence positional error) in the time
between measurements of the separate fringes.  To reduce this effect, 
a fraction of the incoming starlight is redirected to a 
separate beam-combiner.  This beam-combiner is used in a 
``fringe-tracking'' mode \citep{ss80,col99} where it rapidly (10 ms) 
measures the phase of one of the starlight fringes, and adjusts the 
internal delay to keep that phase constant.  
The fringe tracking data is used both in 
real-time as a feed-back servo, after which a 
small residual phase error remains, 
and in post-processing where the measured 
residual error is applied to the data as a feed-forward servo.  
This technique---known 
as phase referencing---has the effect of stabilizing the fringe 
measured by the astrometric beam-combiner.  For this observing mode, 
laser metrology is only required between the two beam combiners through 
the location of the light split (which occurs after the optical delay has been 
introduced), rather than throughout the entire array.  Without phase 
referencing, the astrometric precision obtainable is a factor of a hundred 
worse; see Fig.~\ref{fig:expected}.

To analyze the data, a double fringe 
packet based on Eq.~\ref{double_fringe} is then fit to the data, and the
differential optical path between fringe packets is measured.
A grid in differential right ascension and declination over which to search 
is constructed.  For each point in the search grid the expected differential 
delay is calculated based on the interferometer location, baseline
geometry, and time of observation for each scan.  A model of a 
double-fringe packet is then calculated and compared to the observed
scan to derive a $\chi^2$ value; this is repeated for each scan,
co-adding all of the $\chi^2$ values associated with that point in the
search grid. The final $\chi^2$ surface as a function of differential
R.A. and declination is thus derived. The best-fit astrometric
position is found at the minimum-$\chi^2$ position, with uncertainties
defined by the appropriate $\chi^2$ contour---which depends on the
number of degrees of freedom in the problem and the value of the
$\chi^2$-minimum. The final product is a measurement of the apparent 
vector between the stars and associated uncertainty ellipse.  
Because the data were obtained with a single-baseline
instrument, the resulting error contours are very elliptical, with
aspect ratios that sometimes exceed 10:1.

\begin{figure}[h]
   \centerline{\includegraphics[height=2.5in]{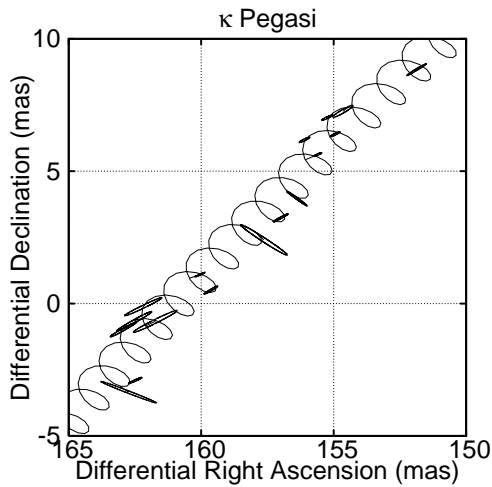}}
   \caption[Visual orbit of $\kappa$ Pegasi]
    { \label{tripleOrbits}
      The visual orbits of $\kappa$ Pegasi 
      showing perturbations by the third component.  The spiral line 
      represents the apparent motion of the short-period pair's center 
      of light; the ellipses represent the $1\sigma$ uncertainties for 
      PHASES measurements.
    }
\end{figure}

\begin{figure}[h]
   \centerline{\includegraphics[width=6.5cm]{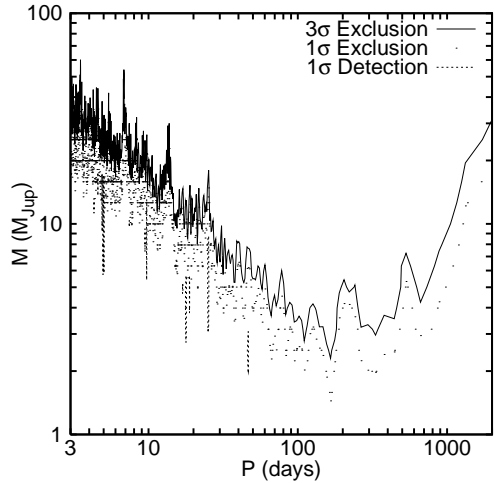}}
   \caption[13 Pegasi Mass-Period Companion Phase Space]
{ \label{fig:207652_phase_space}
The 13 Pegasi Mass-Period companion phase space shows PHASES observations 
can rule out tertiary objects as small as two Jupiter masses.  A few 
mass-period combinations introduce slight improvements over the 
single-Keplerian model, but none of these are more significant than 
$1.7\sigma$, and are probably not astrophysical in origin.  There is a long 
period cutoff in sensitivity due to the finite span of the observations.  
Similar detection 
limits have recently been published on other PHASES targets 
\citep{Mute06Limits}.
}
\end{figure}

The Palomar High-precision Astrometric Search for Exoplanet Systems
(PHASES) program uses this technique to monitor $\sim 50$ binaries to 
search for substellar companions.  
$\kappa$ Pegasi is a well-known, nearby triple star system.  It consists of 
a ``wide'' pair with semi-major axis 235 mas (8.14 $\AU$), one component 
of which is a single-line spectroscopic binary (semi-major axis 
2.5 mas, physical separation 0.087 $\AU$; Fig.~\ref{tripleOrbits}).  
The perturbation due to the unseen (faint) 
short-period component is evident; similar sized perturbations with longer 
orbital periods would indicate the presence of planetary companions.  Figure 
\ref{fig:207652_phase_space} shows the mass-period phase space in 
which PHASES observations show companions do not exist in face-on, circular 
orbits in the 13 Pegasi system.

\subsubsection{Radial Velocities}\label{SRV}


In the case when a composite spectrum of a binary star is observed, the 
classical approach with the iodine cell (described in section \ref{wideRV}) 
cannot be used since it is 
not possible to observationally obtain two separate template spectra 
of the binary components. To resolve this problem, one can proceed as 
follows. First, one always takes two sequential exposures of each (binary) 
target---one with and the other without the cell.  This is contrary to 
the standard approach for single stars where an exposure without the 
cell (a template) is taken only once. This way one obtains an instantaneous 
template that is used to model only the adjacent exposure taken with the 
cell. Next, one performs the usual least-squares fit and obtain the parameters 
described in eq.~\ref{i2::}.  Obviously, 
the derived Doppler shift, $\Delta\lambda_i$  
(where $i$ denotes the epoch of the observation), 
carries no meaning since each 
time a different template is used.  Moreover, it describes a Doppler "shift" 
of a composed spectrum that is typically different at each epoch. 
However, the parameters---in particular the wavelength solution and the 
parameters describing PSF---are accurately determined and can be used to 
extract the star spectrum, $I^{\star,i}_{obs}(\lambda)$, for each epoch $i$, 
by inverting the eq.~\ref{i2::}:  
\begin{equation}
\label{met::}
I^{\star,i}_{obs}(\lambda) = [I^{i}_{obs}(\lambda)\,\otimes^{-1}\,PSF^{i}]
/T_{I_{2}}(\lambda), 
\end{equation}
where $\otimes^{-1}$ denotes deconvolution, and $PSF^{i}$ represents the
set of parameters describing PSF at the epoch $i$. Such a star spectrum has
an accurate wavelength solution, is free of the I$_2$ lines and the influence
of a varying PSF. In the final step, the velocities of both components
of a binary target can be measured with the well know two-dimensional
cross-correlation technique TODCOR \citep{Zucker1994} using as templates the
synthetic spectra derived with the ATLAS~9 and 
ATLAS~12 programs \citep{Kurucz1995} 
and matched to the observed spectrum, $I_{s}(\lambda)$. The formal errors of 
the velocities can be derived from the scatter between the velocities from 
different echelle orders or using the formalism of TODCOR \citep{Zucker1994}. 
The technique currently produces RVs of binary stars with an average precision 
of 20 ${\rm m \, s^{-1}}$ \citep{Konacki04}. Improvements to the 
technique are being introduced to reach the level below 10 ${\rm m \, s^{-1}}$.

With the iodine technique \cite{Konacki04} has initiated the first RV survey
for circumprimary or circumsecondary planets of binary or
multiple stars (mainly hierarchical triples). The survey's sample
of $\sim 450$ binaries (northern and southern hemisphere) was 
selected based on the following criteria. (1) The apparent separation 
of the components had to be smaller than the width of the slit (e.g., 0.6 
arcseconds for Keck-I/HIRES) to avoid possible systematic effects. 
Such systems will remain unresolved under most seeing conditions. 
(2) The brightness ratio between the components should not be too large 
(at the order of 10 or less) to be able to clearly identify spectra of 
both components. (3) The orbits of the binaries should be well known to 
constitute a firm ground on which one can discuss possible detection (or lack) 
of planets and substellar companions in the context of the binary 
characteristics. All these requirements could be satisfied by targeting 
a subset of speckle binaries that have determined orbits from the Catalog 
of Orbits of Visual Binary Stars \citep{hart01}, containing 1700 
binaries, of which 1300 have projected semi-major axes smaller than 1 
arcsecond. The sample is sufficiently large to produce meaningful statistics. 
Also, such binaries have been ignored by previous RV studies.

The survey was initiated in 2003 at the Keck-I/HIRES 
and so far has covered about 50$\%$ of the northern hemisphere 
target sample, of which $\sim$150 stars have been observed at least 
twice. In July 2005, the first candidate planet in a triple 
star system HD 188753 was announced \citep{Konacki2005}. A few more 
candidates as well as HD 188753 are currently being scrutinized. In 
the course of the Keck survey, a dozen of new triple star systems 
with the third bodies in the low-mass star or even brown 
dwarf regime have been also identified and will be published in the 
near future.

%


\subsubsection{Eclipse Timing}

Should a binary happen to be oriented with its orbital plane in the line 
of sight, it will exhibit eclipses as one star passes in front of another.  
For a binary with separation wide enough to allow for stable planetary systems 
to exist around just one component, the probability of such an alignment is 
extremely small, and very few targets are accessible.  However, should one 
find such a fortunate happenstance, one can detect the planetary companions 
by precision timing of the eclipses.  Clearly, this method has no direct 
analog for single systems.

This method can detect S-type planets or similarly moons of 
transiting planets.  For this evaluation it is 
assumed that the depth of the planet (or moon) eclipse is sufficiently small 
as to be ignored (in such a detection, one could then reevaluate light curves 
to look for such transit signals) and the binary orbit is circular.  The 
velocity of the binary orbit and the offset of the star-planet 
CM from that of the star by itself determines the timing variation observed 
as 
\begin{eqnarray}
\Delta t & = & x_{CM}/v_b \nonumber\\
         & = & \left(\frac{a_p \sin \phi M_p}{M_2+M_p}\right)\left(\frac{P_b}{2\pi a_b}\right)\nonumber\\
         & \approx & 57\,{\rm seconds}\times\left(P_b/{\rm month}\right)\frac{a_p}{\left( a_b/7 \right) }\frac{\left( M_p/M_\Jupiter \right) }{\left( M_2/\Msun \right) } \sin \phi
\end{eqnarray}
\noindent where $M_2$ is the mass of star 2 (or the transiting 
planet, assumed to host the S-type companion), 
$M_p$ is the mass of the S-type object orbiting $M_2$, 
$\phi$ is angle between the planet's orbital angular momentum vector and the 
direction of motion of the host star during eclipse, and 
$P_b$ is the period of the binary orbit.  
The timing delays due to orbit of $M_2$ about the $M_2$-$M_p$ CM.  The factor 
of 7 appears in the final form is an approximate criteria for stability, that the planet semimajor axis is 7 times smaller than that of the binary
(this factor varies by the system, and can be determined through 
detailed simulations); the above is thus an upper limit for the timing 
effect.  Converting the se mimajor axis to orbital periods, 
\begin{eqnarray}\label{STimingPert}
\Delta t & \approx & 41\,{\rm seconds}\times\left(P_b/{\rm month}\right)^{\frac{1}{3}}\left(P_p/{\rm day}\right)^{\frac{2}{3}}
                                      \frac{M_p/M_\Jupiter}{\left(M_b^{\frac{1}{3}}M_2^{\frac{2}{3}}\right)/\Msun} \sin \phi \nonumber\\
         & \approx & 65\,{\rm seconds}\times\left(P_b/{\rm month}\right)^{\frac{1}{3}}\left(P_p/{\rm day}\right)^{\frac{2}{3}}
                                      \frac{M_p / M_\Jupiter}{M_b/\Msun} \sin \phi \nonumber\\
\end{eqnarray}
\noindent where $M_b = M_1 + M_2 + M_p$.  
The final form assumes $M_1 \approx M_2$, 
in which case the maximum stable planet period is a thirteenth that 
of the binary period, implying days and months are the natural units for each 
respectively (S-type planets cannot exist in much shorter period systems, and 
longer period systems are even less likely to show eclipses).  The equivalent 
relationship for a moon orbiting an eclipsing Jupiter is
\begin{equation}
\Delta t \approx 13.3\,{\rm seconds}\times\left(P_b/{\rm month}\right)^{\frac{1}{3}}\left(P_p/{\rm day}\right)^{\frac{2}{3}}
				        \frac{M_p / M_\oplus}{\left( M_b / \Msun \right)^{\frac{1}{3}}\left( M_2 / M_\Jupiter \right)^{\frac{2}{3}}} \sin \phi
\end{equation}
\noindent where now the $b$ subscript refers to the star-Jupiter 
analog system and $p$ to the Jupiter analog's moon.

The precision with which eclipse minima can be timed is derived using
standard $\chi^2$ fitting techniques.  Assume a photometric data set
$\{ y_i\}$ occurring at times $\{t_i\}$ with measurement precisions
$\{\sigma_{i}\}$, and a model photometric light curve of flux
$F(t-t_0)$.  The corresponding intensity is 
$I(t-t_0) = f F(t-t_0) \pi D^2\Delta t / 4$, 
where $f$ ($0 \le f \le 1$) is the fractional efficiency and
throughput of the telescope, $D$ is the telescope diameter, and
$\Delta t$ is the sample integration time.  ($F(t-t_0)$ might be
determined to high precision by observing multiple eclipse events.)
The fit parameter $t_0$ is uncertain by an amount equal to the
difference between the value for which $\chi^2$ is minimized and that
for which it is increased by one:  
$1 + \chi^2(t_0)  = \chi^2(t_0 + \sigma_{t_0})$, 
\begin{eqnarray*}
1 + \sum_{i=1}^{N}\left[\frac{y_{i} - I\left( t_i - t_0 \right)}{\sigma_{i}} \right]^2 
& = & \sum_{i=1}^{N}\left[\frac{y_{i} - I\left( t_i - t_0 - \sigma_{t_0} \right)}{\sigma_{i}} \right]^2 \\
& \approx & \sum_{i=1}^{N}\left[
\frac{y_{i} - I\left( t_i - t_0\right) 
- \left( \frac{\partial I\left( t \right)}{\partial t}\right)_{t_i-t_0} \sigma_{t_0}}
{\sigma_{i}} 
\right]^2\nopagebreak
\end{eqnarray*}\nopagebreak
\begin{small}
\begin{displaymath}\nopagebreak
 =  \sum_{i=1}^{N}
\left( 
\left[ \frac{y_{i} - I\left( t_i - t_0 \right)}{\sigma_{i}} \right]^2 + 
\left[ \frac{\left( \frac{\partial I\left( t \right)}{\partial t}\right)_{t=t_i-t_0} \sigma_{t_0}}{\sigma_{i}} \right]^2 
- 2\frac{\left( \frac{\partial I\left( t \right)}{\partial t}\right)_{t=t_i-t_0}
\left[ y_{i}  - I\left( t_i - t_0 \right) \right] \sigma_{t_0} }{\sigma_{i}^2}
\right).
\end{displaymath}\nopagebreak
\end{small}\nopagebreak
\noindent Because $t_0$ is the minimizing point, the first derivative
of $\chi^2$ at $t_0$ is zero, giving 
\begin{small}
\begin{equation}
\left( \frac{\partial \chi^2\left(t \right)}{\partial t}\right)_{t=t_0} =
2 \sum_{i=1}^{N}
\frac{\left( \frac{\partial I\left( t \right)}{\partial t}\right)_{t=t_i-t_0} \left( y_{i} - I\left( t_i - t_0 \right) \right) \sigma_{t_0} }{\sigma_{i}^2} = 0.
\end{equation}
\end{small}
\noindent Rearrangement of terms leads to
\begin{equation}\label{lightcurve_timing_formula}
\sigma_{t_0} = \frac{1}{\sqrt{\sum_{i=1}^{N}\left( \frac{\left( \frac{\partial I\left( t \right)}{\partial t}\right)_{t=t_i-t_0}}{\sigma_{i}} \right)^2}} \nonumber \approx \frac{\sigma_I}{\sqrt{\sum_{i=1}^{N} \left( \frac{\partial I\left( t \right)}{\partial t}\right)_{t=t_i-t_0}^2}}.
\end{equation}

\begin{figure}[tbp]
   \centerline{\includegraphics[height=7cm]{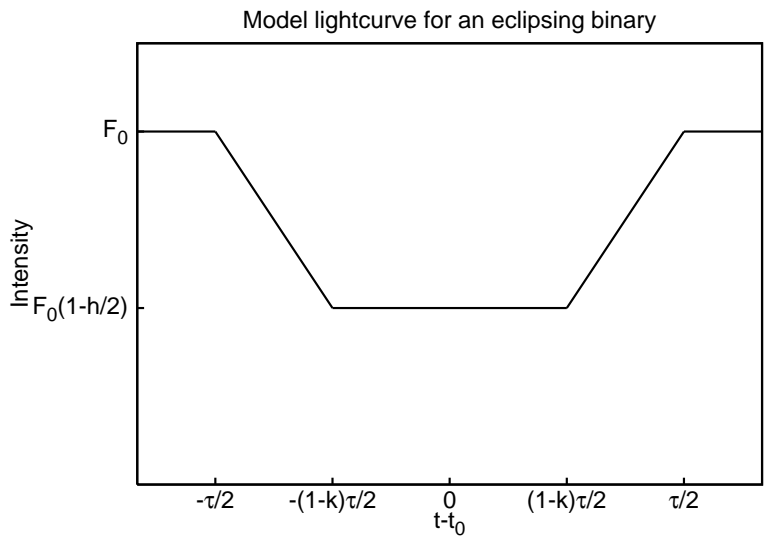}}
   \caption[Eclipsing Binary Model Light Curve] 
{ \label{fig:eclipseModel}
Eclipsing binary model light curve.
}
\end{figure}

An eclipse of length $\tau$ is approximated as a trapezoid-shape light
curve (see Figure \ref{fig:eclipseModel}) 
with maximum and minimum photon fluxes $F_0$ and $F_0(1-h/2)$
($h$ is a dimensionless positive number producing an eclipse depth of
$hF_0/2$; in the case of a faint secondary, $h$ is roughly twice the
ratio of the squares of the stellar radii, $2 R_2^2/R_1^2$).  The
ingress and egress are each assumed to be of length $k\tau/2$
($k\approx 2 R_2/(R_1+R_2)$ is unity in the case of an eclipsing
binary with components of equal size, when the trapezoid becomes a
``V''-shape).  
In functional form, this model is: 
\begin{equation}
F(t-t_0) = \left\{ \begin{array}{lrcl}
F_0 & & t-t_0 & \le -\tau/2 \\
F_0\left(1 - ht/\left(k \tau \right) - h/\left(2 k \right) \right) & - \tau/2 \le & t-t_0 & \le -\tau/2 + k\tau/2 \\
F_0\left(1 - h/2\right) & -\tau/2 + k\tau/2 \le & t-t_0 & \le \tau/2 - k\tau/2 \\
F_0\left(1 + ht/\left(k \tau \right) - h/\left(2 k \right) \right) & \tau/2 - k\tau/2 \le & t-t_0 & \le \tau/2 \\
F_0 & \tau/2 \le & t-t_0 & 
\end{array}
\right.
\end{equation}
\noindent (also, see Figure \ref{fig:eclipseModel}).

Only the portions of the light curve during ingress and
egress have nonzero $\frac{\partial F\left( t \right)}{\partial t}$
(in more accurate models, the light curve slope will be nonzero but
small in other regions, and will not contribute much to the sum); this
slope is 
$\left| \frac{\partial F\left( t \right)}{\partial t} \right| 
= h F_0/\left(k \tau \right)$.  

The number of data points contributing to the sum is thus $N = g k
\tau / \Delta t$, where $0 \le g \le 1$ is the fraction of the eclipse
observed (and also accounts for the fraction of time lost to, e.g., camera
readout) and $\Delta t$ is the integration time for each measurement.

The measurement noise $\sigma_I$ is given by
\begin{equation}\label{photometryNoiseEquation}
\sigma_I = 
\left(
I + \sigma^2_{sc} + I_{bg} + n_{dark} \Delta t + \sigma^2_{rn}
\right)^{\frac{1}{2}}
\end{equation}
\noindent where $I_{bg} = f F_{bg} \pi D^2 \Delta t/4$ 
is is the sky background, $n_{dark}$ is detecter dark current, 
$\sigma_{rn}$ is detector read noise, and $\sigma_{sc}$ is 
scintillation noise given by \cite{Young1967} as 
\begin{eqnarray}
\sigma_{sc} & = & 0.09 I \left(D/1\,{\rm cm}\right)^{-2/3} X e^{-h/\left(8000\,{\rm m}\right)} / \left(2 \Delta t / 1\,{\rm second}\right)^{\frac{1}{2}} \\
            & \approx & 0.003 I \left(D/1\,{\rm m}\right)^{-2/3} / \left(\Delta t /1\,{\rm second}\right)^{\frac{1}{2}}
\end{eqnarray}
\noindent where $X$ is the airmass and $h$ is the altitude of the 
observatory.  The drop in noise during eclipse is ignored (a factor less 
than $\approx 1.4$) and equation \ref{photometryNoiseEquation} is combined 
with equation \ref{lightcurve_timing_formula} to obtain an overall 
timing precision (in seconds) of
\begin{scriptsize}
\begin{eqnarray}\label{timing_prec_2}
\sigma_{t_0} & = & \sqrt{\frac{k \left( \tau / 1\,{\rm second}\right)}{gh^2}}
                   \left(
                         \frac{4}{f F_0 \pi \left(D/1\,{\rm m}\right)^2} + 
                         \frac{9\times10^{-6}}{\left(D/1\,{\rm m}\right)^{4/3}} + 
                         \frac{\pi D^2 f F_{bg}/4 + n_{dark} + \sigma_{rn}^2/\left( \Delta t / 1\,{\rm second}\right)}
                         {f^2 F_0^2 \pi^2 \left(D/1\,{\rm m}\right)^4/16}
                  \right)^{\frac{1}{2}}\nonumber\\
       & \approx & 0.18\,{\rm seconds}\times \sqrt{\frac{k \left(\tau/1\,{\rm hr}\right)}{fgh^2}}
                   \left(\frac{10^{\left(V-12\right)/2.5}}{\left(D/1\,{\rm m}\right)^2} + \frac{f}{\left(D/1\,{\rm m}\right)^{4/3}}\right)^{\frac{1}{2}}.
\end{eqnarray}
\end{scriptsize}
\noindent The final term in the first line (associated with dark current, read 
noise, and background) is generally smallest and will be ignored.  
In most cases, the second term---associated with scintillation---is dominant 
(though zero in the case of space-based observatories).  The exponent of 
$(V-12)/2.5$ shows that for meter-sized telescopes, photon noise is only 
dominant for stars fainter than twelfth magnitude.

Systematic and astrophysical noise sources may have effects that limit 
the actual precisions achieved.
Mass transfer between stars can cause drifts in orbital 
periods.  Variations of this type are non-periodic, distinguishing 
them from companion signals.  
\cite{Applegate} has shown that gravitational coupling 
to the shapes of magnetically active stars can cause periodic 
modulations over decade timescales.  This mechanism 
requires the star to be inherently variable; false positives 
can be removed using the overall calibrations of photometric 
data.  It is possible that star spots will have large effects 
on timing residuals that are particularly difficult to 
calibrate \citep{Starspots}.  
Due to orbit-rotation tidal locking, the effect of a starspot 
on the light curve can be detected from the light curves of several orbits, 
and starspot fitting potentially can remove the timing biases introduced.  

The timescale for eclipses of such long period binaries is of the order 
of 12 hours.  Comparing eqs.~\ref{STimingPert} 
and \ref{timing_prec_2}, one finds that S-type 
planets with periods of a few days can be detected around either star in 
eclipsing binaries with month-long periods if they mass more than
\begin{equation}
M_p/M_\earth \gtrsim 3 
 \frac{\left(M_b/\Msun\right)}
      {\left(P_b/{\rm month}\right)^{\frac{1}{3}}\left(P_p/{\rm day}\right)^{\frac{2}{3}}} 
 \sqrt{\frac{k \left(\tau/12\,{\rm hr}\right)}
      {fgh^2}}
 \left(\frac{10^{\left(V-12\right)/2.5}}{\left(D/1\,{\rm m}\right)^2} + \frac{f}{\left(D/1\,{\rm m}\right)^{4/3}}\right)^{\frac{1}{2}}.
\end{equation}

Eq.~\ref{timing_prec_2} indicates that a meter-class ground-based 
telescope can time a giant planet transit ($h \approx 0.02$, $k \approx 0.18$) 
to approximately 9.4 seconds in the regime where the photometric precision is 
dominated by scintillation noise, assuming a Jupiter sized planet orbiting a 
star of solar size and mass with period of a month (implying 6-hour duration 
eclipses).  This precision is sufficient to find Earth mass moons.
For bright stars, space-based observatories offer even better precisions.  
Unfortunately, no transiting exoplanets with periods this long have yet been 
discovered.

Space-based photometric missions such as Kepler have as their primary goal 
the detection of Earth-like planets via transits of the planet across the 
star.  However, such photometric events can be explained by other 
astrophysical phenomena, such as a transiting Jupiter blended with a 
background star, so these results may be unreliable.  However, 
Earth-like moons of transiting Jupiters might 
be identified through timing, and it is possible to confirm the nature of 
such a system.  In such a scenario, a transiting Jupiter can be positively 
confirmed by ground-based radial velocity observations.  Once this has 
been established, variations in the transit times would be used to detect 
Earth-sized moons.  Because these photometric missions have limited 
lifetimes ($\approx 3$ years), detections of moons are only possible 
for short period (few months or less) Jupiters, for which many transit 
events can be observed (unless a follow-up 
ground-based campaign is pursued with large telescopes).  
If the planet/moon are to be in the habitable zone, 
one must look for such systems around late-type (cool) stars.  It is possible 
that such systems have the greatest likelihood of being habitable; 
tidal-locking of the Earth-sized moon to 
the Jupiter-like planet would ensure that the moon has day/night 
cycles and stabilize its rotational axis similar to the way 
in which the Earth's is stabilized by its own moon.  Both of these 
conditions have been argued as favorable for life 
\citep[see, for example,][]{earthAxisStability}.


\subsubsection{Observational Precisions}

\begin{figure}[tbp]
\begin{center}
\includegraphics[height=3.5in]{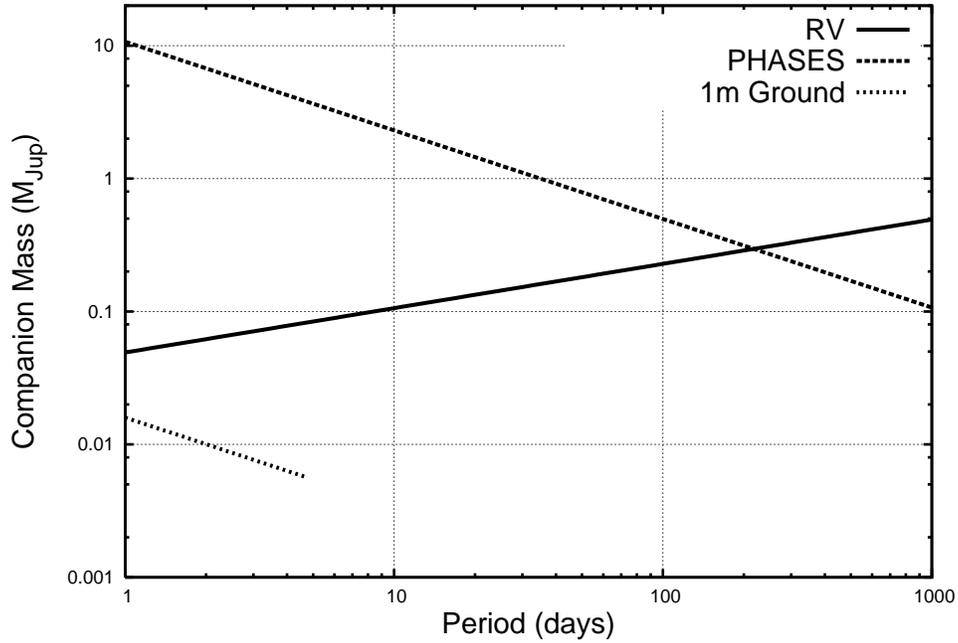}
\end{center}
\caption[Sensitivity to S-Type Planets in Narrow Binaries]
{\label{fig:narrowS} 
Sensitivity to S-Type planets in narrow binaries, comparing 
astrometric, radial velocity, and eclipse timing techniques.  All calculations 
assume solar mass stars.  The PHASES sensitivity assumes 20 $\microas$ 
precision and a distance to the system of 20 pc.  Eclipse timing assumes 
a 1 m photometric telescope observing a $V=10$ magnitude system with 
binary orbital period 2 months (longer period systems are even less likely to 
show eclipses); the eclipse timing sensitivity curve only extends to the 
region where planets are likely to have stable orbits.
}
\end{figure}

Figure \ref{fig:narrowS} shows the companion 
masses one can detect for each method, assuming 
20 $\microas$ ground-based astrometry, $20 {\rm m\, s^{-1}}$ RV, and 1 m
ground-based photometric telescope for eclipse timing.

\section{P-Type (Circumbinary) Planets}

All the confirmed planets found in binary systems thus far are in 
S-type orbits.  Discovery of circumbinary planets would consitute a new class 
of solar system, and would inspire new considerations to the interplay 
between system dynamics and planet formation.


\subsection{Radial Velocities}

A circumbinary planet will exhibit two indirect effects 
on the velocities of the stellar components of the system.
First, the apparent system velocity will vary in a periodic manner due 
to the motion of the binary about the system barycenter.  
Second, the finite speed of light will cause apparent changes in the phase 
of the binary orbit.  These effects may be detectable using modern 
observational techniques.

The first effect is that the binary will exhibit periodic
changes in the apparent system velocity; this is the same
effect as seen in a single star. However, it may be harder
to detect for three reasons: (1) the binary system is usually more
massive than a single star of the same magnitude, (2) extremely 
short-period planet orbits (to which system velocity measurements are most 
sensitive) are unstable around binaries, and (3) the presence of two 
sets of spectral lines may complicate the measurement, as in section 
\ref{SRV}.  Equation \ref{vel_reflex} shows that a Jupiter massed 
planet with the shortest period stable orbit around a 10-day period binary 
causes a $2 \Msun$ binary to move about its barycenter by 
$\approx 40\,{\rm m\,s^{-1}}$, with the amplitude decreasing 
as the square root of planet orbit semimajor axis.  
Radial velocity observations with the $20\,{\rm m\,s^{-1}}$ precision 
demonstrated with Konacki's method can detect Jupiter-like 
planets in orbits of size $\approx 4$ \AU~or less, 
down to the critically stable orbit.

The second observable effect is the additional light travel time as
the binary system undergoes reflex motion caused by the planet.
The magnitude of this effect is given by 
\begin{equation}\label{time_reflex}
\Delta t = 2 \frac{ a_p M_p \sin i_p}{ c M_b }\nonumber\\
         = 0.95\,{\rm seconds} \times \frac{ \left(a_p/1\,\rm{AU}\right) 
                           \left(M_p/M_\Jupiter\right) \sin i_p}{ M_b/\Msun }.
\end{equation}

Following a similar derivation as that for finding the 
expected precision of eclipse timing, 
one finds the precision with which one can estimate the orbital phase
of a binary based on radial velocity measurements is
\begin{equation}
\sigma_{\phi} = \frac{\sigma_{rv}}{\sqrt{\sum_i 
(\frac{\partial v_i}{\partial \phi})^2}}, 
\end{equation}
\noindent where $\sigma_{rv}$ is the radial velocity measurement precision 
and $\frac{\partial v_i}{\partial \phi}$ is the derivative of the model
radial velocity curve with respect to orbital phase, evaluated
at times $t_i$.  The timing precision corresponding to the 
phase precision derived is given by 
$\frac{\sigma_{\phi}}{2\pi} = \frac{\sigma_t}{P_b}$.

Approximating the binary orbit as circular, 
$v\left( t \right) \approx K\cos\left(\frac{2\pi t}{P_b}+\phi\right)$.
If $N$ measurements (each with two measured velocities, one for each star) 
are approximately evenly distributed in phase, 
\begin{eqnarray}
\sigma_{\phi} & = & \frac{\sqrt{2}\sigma_{rv}}{\sqrt{\left(2N-12\right)} K}\\
\sigma_t & = & \frac{P_b\sigma_{rv}}{\sqrt{2\left(2N-12\right)}\pi K}, 
\end{eqnarray}
\noindent where 12 is the number of degrees of freedom for the model.

If the lines from both stars are observed, the effective $K$ is $K1+K2$ and 
the resulting ($1\sigma$) minimum detectable mass is thus
\begin{small}
\begin{equation}
M_p = 41.4 M_\Jupiter \times \frac{\left( \sigma_{rv}/{\rm 20\,m\,s^{-1}}\right) \left(P_b/{\rm 10\,days}\right)^{4/3} \left(M_b/\Msun\right)^{2/3}}
{\sqrt{2N-12}\sin{i_b}\sin{i_p}\left( a_p/{\rm 1\,AU}\right)} , 
\end{equation}
\end{small}
\noindent where $i_b$ and $i_p$ are the inclinations 
of the binary and planet orbits, respectively.  Twenty-five 
$20\,{\rm m\,s^{-1}}$ radial velocity measurements of the 
``prototypical'' system could detect moderate-mass brown 
dwarfs ($\approx 30 M_\Jupiter$) at critical orbit.
Objects at the planet/brown dwarf 
threshold of 13 $M_\Jupiter$ are only detectable in orbits larger than 
0.82 AU around a ten-day binary of sunlike stars.
Alternatively, if only one set of lines are observed, the resulting expression is
\begin{small}
\begin{equation}
M_p = 41.4 M_\Jupiter \times \left(1+\frac{M_1}{M_2}\right)\frac{\left( \sigma_{rv}/{\rm 20\,m\,s^{-1}}\right) \left(P_b/{\rm 10\,days}\right)^{4/3} \left(M_b/\Msun\right)^{2/3}}
{\sqrt{N-11}\sin{i_b}\sin{i_p}\left( a_p/{\rm 1\,AU}\right)}, 
\end{equation}
\end{small}
\noindent where $M_1$ is the mass of the star whose lines 
are observed, and $M_2$ is that of the faint star.

High precision radial velocity observations are only possible 
on slowly rotating ($v \sin i < 10\,{\rm m\,s^{-1}}$) stars; measurements of 
more rapidly rotating stars are limited by line broadening to 
levels worse than the nominal $20\,{\rm m\,s^{-1}}$ that 
has been referenced by this work.  
This effect is particularly important for finding planets around 
short-period binaries, in which the stars' rotation rates are often tidally 
locked to the binary orbital period; these rotation rates limit the observed 
precisions for systems with periods approximately five days or less.

\begin{figure}[tbph]
   \centerline{\includegraphics[height=7cm]{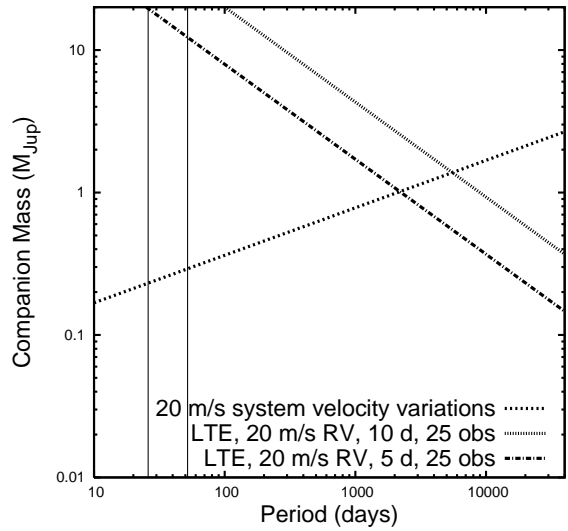}}
   \caption[Radial Velocity Sensitivity] 
{ \label{fig:rvPhase}
Sensitivity of radial velocity measurements to circumbinary planets.  
The two vertical lines at the left represent the approximate 
critical orbits around 5-day (to the left) and 10-day period 
binaries; shorter period companions have unstable orbits.  
Stars whose rotation rates are tidally locked to 
orbital periods less than about 5 days show sufficient rotational line 
broadening to prevent $20\,{\rm m\,s^{-1}}$ radial velocity precisions.
The calculations assume the binary consists of two 
stars each massing 1 $\Msun$.
}
\end{figure}


\subsection{Eclipse Timing}

It has long been recognized that periodic shifts in the observed times
of photometric minima of eclipsing binaries can indicate the presence
of an additional component to the system (see, for example, 
\cite{woltjer1922, irwin1952, Frieboes-Conde, Doyle98}).  The
amplitude of the effect is given by eq.~\ref{time_reflex}.  As with RV
measurements, there is a mass/inclination ambiguity; the following
derivation assumes no correlation between binary and planet inclinations.

Dividing the precision of an individual measurement by $N_{obs} - 6$
(where $N_{obs}$ is the number of eclipses observed and there are 6 parameters 
to a timing perturbation fit, two periods and the eccentricity, angle of periastron, 
epoch of periastron, and mass ratio of the wide pair), converting $F_0$ to $V$ magnitude, 
and combining eq.~\ref{timing_prec_2} 
with that for the timing effect of reflex motion (eq.~\ref{time_reflex})
gives a minimum detectable companion mass of
\begin{equation}
M_p = 0.19 M_\Jupiter \times \sqrt{\frac{k \left(\tau/1\,{\rm hr}\right)}{fgh^2 \left(N_{obs}-6\right)}}
\frac{M_b/\Msun}{\left( a_p/1\,{\rm AU} \right) \sin i_p}
\left(\frac{10^{\left(V-12\right)/2.5}}{\left(D/1\,{\rm m}\right)^2} + \frac{f}{\left(D/1\,{\rm m}\right)^{4/3}}\right)^{\frac{1}{2}}.
\end{equation}

\begin{figure}[tbph]
   \centerline{\includegraphics[height=7cm]{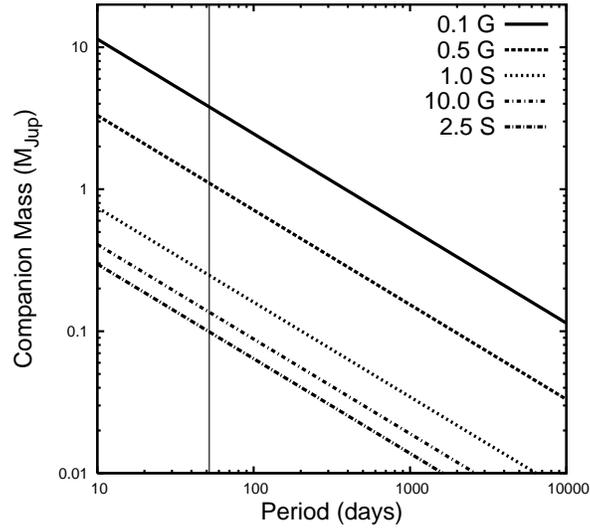}}
   \caption[Eclipse Timing Sensitivity] 
{ \label{fig:eclipsePhase}
Sensitivity of eclipse timing measurements to circumbinary planets.  
The vertical line represents the approximate critical orbit around 
a 10-day period binary.  The calculations assume the binary consists 
of two stars each massing 1 $\Msun$, 6 hour eclipses, 
$N_{obs} = 25$ observations (150 total hours of data), 
$V=10$ magnitudes, and 1-$\sigma$ detections.  From top to bottom, 
lines show sensitivity for $D=0.1$ m on the ground, $D=0.5$ m on the ground, 
$D=1.0$ m in space (i.e.~Kepler), $D=10$ m on the ground, and $D=2.5$ m in space (HST, SOFIA).
}
\end{figure}

One might also inquire about the sensitivity of this technique to 
outer planets in systems comprised of a single star and a transiting 
``hot'' Jupiter.  In this case, 
$h \approx 2 R_p^2/R_{star}^2 \approx 0.02$ and 
$k \approx 2 R_p/(R_{star}+R_p) \approx 0.18$, the 
``binary'' is half as massive, 
and the eclipse duration is half as long.  
The companion sensitivity drops by a factor of 8, and the technique 
is (barely) in the range of detecting additional companions of planet mass.  
However, for the typically $V=10$ magnitude transiting planet systems 
being discovered, 3 ${\rm m\,s^{-1}}$ radial 
velocity observations are more sensitive 
than half-meter telescope transit timing for companions with periods up 
to 60 years; even for observatories such as HST and SOFIA (for which 
scintillation noise is small or zero), this transition occurs at 
15 year period companions.

It should be noted that the above description does not account for the 
possibility of resonant orbits, for which timing perturbations can be greatly 
enhanced by many-body dynamics---we have assumed independent Keplerian orbits 
for the subsystems.  Resonant effects on timing perturbations have been 
considered by \cite{holman2005, Agol2005}.




\subsection{Observational Precisions}

\begin{figure}[tbp]
\begin{center}
\includegraphics[height=3.5in]{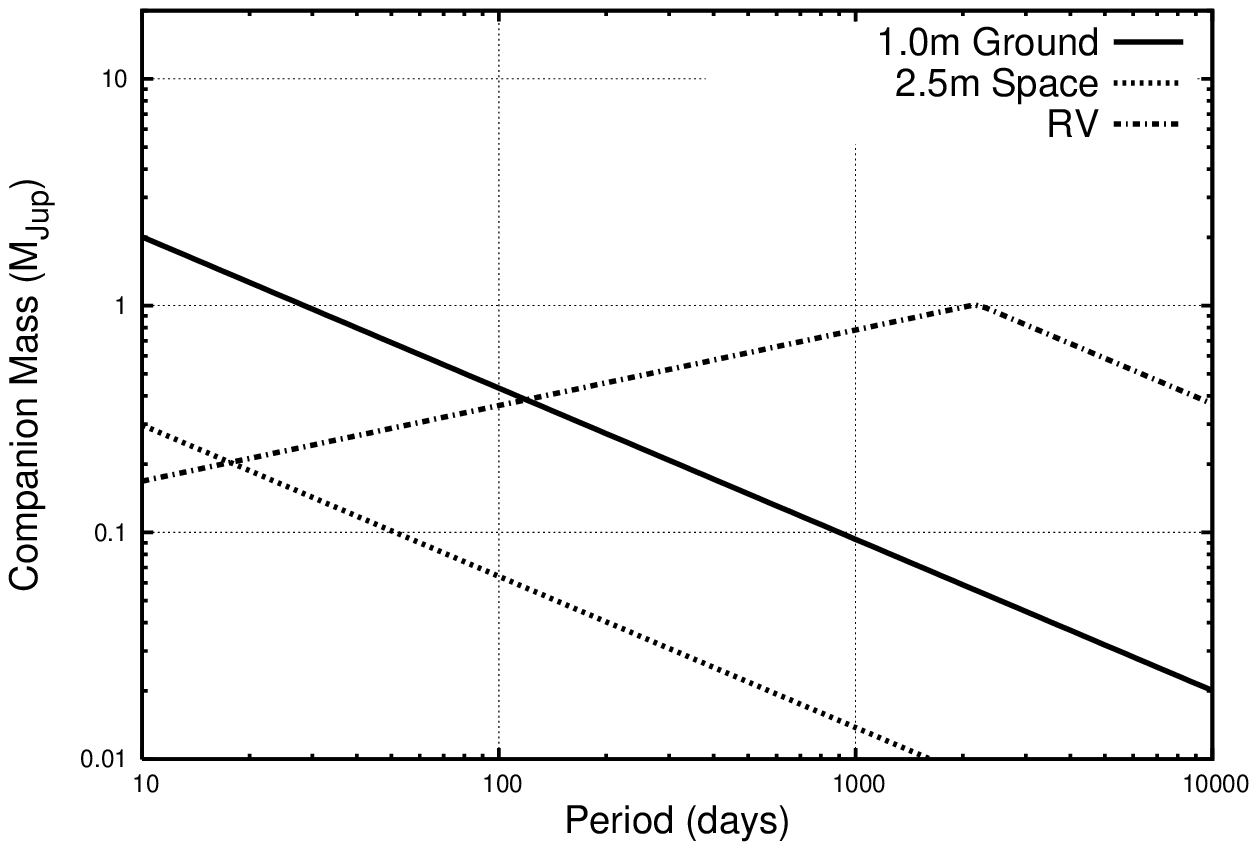}
\end{center}
\caption[Sensitivity to Circumbinary Planets]
{\label{fig:PType} 
Sensitivity to circumbinary planets, comparing 
radial velocity and eclipse timing techniques.  All calculations 
assume solar mass stars.  RV assumes 20 ${\rm m\, s^{-1}}$ precision; 
both the system velocity and apparent period variation observables are 
included for the sensitivity curve, the latter assumes a 5 day binary 
orbital period.  Eclipse timing assumes either 
a 1 m ground-based photometric telescope or 2.5 m space-based 
telescope (such as HST or SOFIA), observing a $V=10$ magnitude system, 
and 25 observations.
}
\end{figure}

Figure \ref{fig:PType} compares the sensitivity of RV and eclipse timing 
to circumbinary planets.

\bibliography{main}

\newcommand{\noopsort}[1]{} \newcommand{\printfirst}[2]{#1}
  \newcommand{\singleletter}[1]{#1} \newcommand{\switchargs}[2]{#2#1}
\begin{thebibliography}{56}
\providecommand{\natexlab}[1]{#1}
\providecommand{\url}[1]{\texttt{#1}}
\expandafter\ifx\csname urlstyle\endcsname\relax
  \providecommand{\doi}[1]{doi: #1}\else
  \providecommand{\doi}{doi: \begingroup \urlstyle{rm}\Url}\fi

\bibitem[{Agol} et~al.(2005){Agol}, {Steffen}, {Sari}, and
  {Clarkson}]{Agol2005}
E.~{Agol}, J.~{Steffen}, R.~{Sari}, and W.~{Clarkson}.
\newblock {On detecting terrestrial planets with timing of giant planet
  transits}.
\newblock \emph{\mnras}, 359:\penalty0 567--579, May 2005.
\newblock \doi{10.1111/j.1365-2966.2005.08922.x}.

\bibitem[{Applegate}(1992)]{Applegate}
J.~H. {Applegate}.
\newblock {A mechanism for orbital period modulation in close binaries}.
\newblock \emph{\apj}, 385:\penalty0 621--629, February 1992.

\bibitem[{Benest}(1988)]{Benest1988}
D.~{Benest}.
\newblock {Planetary orbits in the elliptic restricted problem. I - The Alpha
  Centauri system}.
\newblock \emph{\aap}, 206:\penalty0 143--146, November 1988.

\bibitem[{Benest}(1989)]{Benest1989}
D.~{Benest}.
\newblock {Planetary orbits in the elliptic restricted problem. II - The Sirius
  system}.
\newblock \emph{\aap}, 223:\penalty0 361--364, October 1989.

\bibitem[{Benest}(1993)]{Benest1993}
D.~{Benest}.
\newblock {Stable planetary orbits around one component in nearby binary stars.
  II}.
\newblock \emph{Celestial Mechanics and Dynamical Astronomy}, 56:\penalty0
  45--50, June 1993.

\bibitem[{Benest}(1996)]{Benest1996}
D.~{Benest}.
\newblock {Planetary orbits in the elliptic restricted problem. III. The
  {$\eta$} Coronae Borealis system.}
\newblock \emph{\aap}, 314:\penalty0 983--988, October 1996.

\bibitem[{Benest}(2003)]{Benest2003}
D.~{Benest}.
\newblock {Planetary orbits in the elliptic restricted problem. V.. The ADS
  11060 system}.
\newblock \emph{\aap}, 400:\penalty0 1103--1111, March 2003.

\bibitem[{Boss}(2000)]{Boss2000}
A.~P. {Boss}.
\newblock {Possible Rapid Gas Giant Planet Formation in the Solar Nebula and
  Other Protoplanetary Disks}.
\newblock \emph{\apjl}, 536:\penalty0 L101--L104, June 2000.
\newblock \doi{10.1086/312737}.

\bibitem[{Broucke}(2001)]{Broucke2001}
R.~A. {Broucke}.
\newblock {Stable Orbits of Planets of a Binary Star System in the
  Three-Dimensional Restricted Problem}.
\newblock \emph{Celestial Mechanics and Dynamical Astronomy}, 81:\penalty0
  321--341, December 2001.

\bibitem[{Butler} et~al.(1996){Butler}, {Marcy}, {Williams}, {McCarthy},
  {Dosanjh}, and {Vogt}]{Butler1996}
R.~P. {Butler}, G.~W. {Marcy}, E.~{Williams}, C.~{McCarthy}, P.~{Dosanjh}, and
  S.~S. {Vogt}.
\newblock {Attaining Doppler Precision of 3 M s-1}.
\newblock \emph{\pasp}, 108:\penalty0 500--+, June 1996.

\bibitem[{Campbell} et~al.(1988){Campbell}, {Walker}, and {Yang}]{Campbell1988}
B.~{Campbell}, G.~A.~H. {Walker}, and S.~{Yang}.
\newblock {A search for substellar companions to solar-type stars}.
\newblock \emph{\apj}, 331:\penalty0 902--921, August 1988.
\newblock \doi{10.1086/166608}.

\bibitem[{Chauvin} et~al.(2006){Chauvin}, {Lagrange}, {Udry}, {Fusco},
  {Galland}, {Naef}, {Beuzit}, and {Mayor}]{Chauvin2006b}
G.~{Chauvin}, A.~. {Lagrange}, S.~{Udry}, T.~{Fusco}, F.~{Galland}, D.~{Naef},
  J.~. {Beuzit}, and M.~{Mayor}.
\newblock {Probing long-period companions to planetary hosts. VLT and CFHT near
  infrared coronographic imaging surveys}.
\newblock \emph{A\&A, in press, astro-ph/0606166}, June 2006.

\bibitem[{Colavita}(1994)]{col94}
M.~M. {Colavita}.
\newblock {Measurement of the Atmospheric Limit to Narrow Angle Interferometric
  Astrometry Using the Mark-III Stellar Interferometer}.
\newblock \emph{\aap}, 283:\penalty0 1027--+, March 1994.

\bibitem[{Colavita} et~al.(1999){Colavita}, {Wallace}, {Hines}, {Gursel},
  {Malbet}, {Palmer}, {Pan}, {Shao}, {Yu}, {Boden}, {Dumont}, {Gubler},
  {Koresko}, {Kulkarni}, {Lane}, {Mobley}, and {van Belle}]{col99}
M.~M. {Colavita}, J.~K. {Wallace}, B.~E. {Hines}, Y.~{Gursel}, F.~{Malbet},
  D.~L. {Palmer}, X.~P. {Pan}, M.~{Shao}, J.~W. {Yu}, A.~F. {Boden}, P.~J.
  {Dumont}, J.~{Gubler}, C.~D. {Koresko}, S.~R. {Kulkarni}, B.~F. {Lane}, D.~W.
  {Mobley}, and G.~T. {van Belle}.
\newblock {The Palomar Testbed Interferometer}.
\newblock \emph{\apj}, 510:\penalty0 505--521, January 1999.

\bibitem[{Desidera} et~al.(2006){Desidera}, {Gratton}, {Claudi}, {Barbieri},
  {Bonanno}, {Bonavita}, {Cosentino}, {Endl}, {Lucatello}, {Martinez
  Fiorenzano}, {Marzari}, and {Scuderi}]{Desidera2006b}
S.~{Desidera}, R.~{Gratton}, R.~{Claudi}, M.~{Barbieri}, G.~{Bonanno},
  M.~{Bonavita}, R.~{Cosentino}, M.~{Endl}, S.~{Lucatello}, A.~F. {Martinez
  Fiorenzano}, F.~{Marzari}, and S.~{Scuderi}.
\newblock {Searching for planets around stars in wide binaries}.
\newblock pages 119--126, February 2006.

\bibitem[{Doyle} et~al.(1998){Doyle}, {Deeg}, {Jenkins}, {Schneider}, {Ninkov},
  {Stone}, {Gotzger}, {Friedman}, {Blue}, and {Doyle}]{Doyle98}
L.~R. {Doyle}, H.~{Deeg}, J.~M. {Jenkins}, J.~{Schneider}, Z.~{Ninkov},
  R.~{Stone}, H.~{Gotzger}, B.~{Friedman}, J.~E. {Blue}, and M.~F. {Doyle}.
\newblock {Detectability of Jupiter-to-Brown-Dwarf-Mass Companions Around Small
  Eclipsing Binary Systems}.
\newblock In \emph{ASP Conf. Ser. 134: Brown Dwarfs and Extrasolar Planets},
  pages 224--+, 1998.

\bibitem[{Dvorak}(1982)]{Dvorak1982}
R.~{Dvorak}.
\newblock {Planetary orbits in double star systems}.
\newblock \emph{Oesterreichische Akademie Wissenschaften Mathematisch
  naturwissenschaftliche Klasse Sitzungsberichte Abteilung}, 191:\penalty0
  423--437, 1982.

\bibitem[{Frieboes-Conde} and {Herczeg}(1973)]{Frieboes-Conde}
H.~{Frieboes-Conde} and T.~{Herczeg}.
\newblock {Period variations of fourteen eclipsing binaries with possible
  light-time effect}.
\newblock \emph{\aaps}, 12:\penalty0 1--+, October 1973.

\bibitem[Hartkopf et~al.(2001)Hartkopf, Mason, and Worley]{hart01}
W.~I. Hartkopf, B.~D. Mason, and C.~E. Worley.
\newblock Sixth catalog of orbits of visual binary stars.
\newblock \emph{http://www.ad.usno.navy.mil/wds/orb6/orb6.html}, 2001.

\bibitem[{Hatzes} et~al.(2003){Hatzes}, {Cochran}, {Endl}, {McArthur},
  {Paulson}, {Walker}, {Campbell}, and {Yang}]{hatzes2003}
A.~P. {Hatzes}, W.~D. {Cochran}, M.~{Endl}, B.~{McArthur}, D.~B. {Paulson},
  G.~A.~H. {Walker}, B.~{Campbell}, and S.~{Yang}.
\newblock {A Planetary Companion to {$\gamma$} Cephei A}.
\newblock \emph{\apj}, 599:\penalty0 1383--1394, December 2003.
\newblock \doi{10.1086/379281}.

\bibitem[{Holman} and {Murray}(2005)]{holman2005}
M.~J. {Holman} and N.~W. {Murray}.
\newblock {The Use of Transit Timing to Detect Terrestrial-Mass Extrasolar
  Planets}.
\newblock \emph{Science}, 307:\penalty0 1288--1291, February 2005.
\newblock \doi{10.1126/science.1107822}.

\bibitem[{Holman} and {Wiegert}(1999)]{holman1999}
M.~J. {Holman} and P.~A. {Wiegert}.
\newblock {Long-Term Stability of Planets in Binary Systems}.
\newblock \emph{\aj}, 117:\penalty0 621--628, January 1999.

\bibitem[{Irwin}(1952)]{irwin1952}
J.~B. {Irwin}.
\newblock {The Determination of a Light-Time Orbit.}
\newblock \emph{\apj}, 116:\penalty0 211--+, July 1952.

\bibitem[{Jang-Condell}(2006)]{Jang-Condell2006}
H.~{Jang-Condell}.
\newblock {Constraints on the Formation of the Planet Around HD188753A}.
\newblock \emph{ApJ, in press, astro-ph/050735}, July 2006.

\bibitem[{Konacki}(2005{\natexlab{a}})]{Konacki04}
M.~{Konacki}.
\newblock {Precision Radial Velocities of Double-lined Spectroscopic Binaries
  with an Iodine Absorption Cell}.
\newblock \emph{\apj}, 626:\penalty0 431--438, June 2005{\natexlab{a}}.
\newblock \doi{10.1086/429880}.

\bibitem[{Konacki}(2005{\natexlab{b}})]{Konacki2005}
M.~{Konacki}.
\newblock {An extrasolar giant planet in a close triple-star system}.
\newblock \emph{\nat}, 436:\penalty0 230--233, July 2005{\natexlab{b}}.
\newblock \doi{10.1038/nature03856}.

\bibitem[{Kurucz}(1995)]{Kurucz1995}
R.~L. {Kurucz}.
\newblock {The Kurucz Smithsonian Atomic and Molecular Database}.
\newblock pages 205--+, 1995.

\bibitem[{Lagrange} et~al.(2006){Lagrange}, {Beust}, {Udry}, {Chauvin}, and
  {Mayor}]{Lagrange2006}
A.~. {Lagrange}, H.~{Beust}, S.~{Udry}, G.~{Chauvin}, and M.~{Mayor}.
\newblock {New constrains on Gliese 86 B}.
\newblock \emph{A\&A, in press, astro-ph/0606167}, June 2006.

\bibitem[{Lane} and {Muterspaugh}(2004)]{LaneMute2004a}
B.~F. {Lane} and M.~W. {Muterspaugh}.
\newblock {Differential Astrometry of Subarcsecond Scale Binaries at the
  Palomar Testbed Interferometer}.
\newblock \emph{\apj}, 601:\penalty0 1129--1135, February 2004.

\bibitem[{Lane} et~al.(2000){Lane}, {Colavita}, {Boden}, and {Lawson}]{l00}
B.~F. {Lane}, M.~M. {Colavita}, A.~F. {Boden}, and P.~R. {Lawson}.
\newblock {Palomar Testbed Interferometer: update}.
\newblock In \emph{Proc. SPIE Vol. 4006, p. 452-458, Interferometry in Optical
  Astronomy, Pierre J. Lena; Andreas Quirrenbach; Eds.}, pages 452--458, July
  2000.

\bibitem[{Laskar} et~al.(1993){Laskar}, {Joutel}, and
  {Robutel}]{earthAxisStability}
J.~{Laskar}, F.~{Joutel}, and P.~{Robutel}.
\newblock {Stabilization of the earth's obliquity by the moon}.
\newblock \emph{\nat}, 361:\penalty0 615--617, February 1993.

\bibitem[{Lawson}(2000)]{Lawson2000}
P.~R. {Lawson}, editor.
\newblock \emph{{Principles of Long Baseline Stellar Interferometry}}, 2000.

\bibitem[{Lissauer}(1993)]{Liss1993}
J.~J. {Lissauer}.
\newblock {Planet formation}.
\newblock \emph{\araa}, 31:\penalty0 129--174, 1993.
\newblock \doi{10.1146/annurev.aa.31.090193.001021}.

\bibitem[{Marcy} and {Butler}(1992)]{Marcy1992}
G.~W. {Marcy} and R.~P. {Butler}.
\newblock {Precision radial velocities with an iodine absorption cell}.
\newblock \emph{\pasp}, 104:\penalty0 270--277, April 1992.

\bibitem[{Mugrauer} and {Neuh{\"a}user}(2005)]{Mugrauer2005}
M.~{Mugrauer} and R.~{Neuh{\"a}user}.
\newblock {Gl86B: a white dwarf orbits an exoplanet host star}.
\newblock \emph{\mnras}, 361:\penalty0 L15--L19, July 2005.
\newblock \doi{10.1111/j.1745-3933.2005.00055.x}.

\bibitem[{Muterspaugh} et~al.(2006){Muterspaugh}, {Lane}, {Kulkarni}, {Burke},
  {Colavita}, and {Shao}]{Mute06Limits}
M.~W. {Muterspaugh}, B.~F. {Lane}, S.~R. {Kulkarni}, B.~F. {Burke}, M.~M.
  {Colavita}, and M.~{Shao}.
\newblock {Limits to Tertiary Astrometric Companions in Binary Systems.}
\newblock \emph{\apj}, 653, November 2006.
\newblock \doi{10.1086/508743}.

\bibitem[{Nelson}(2000)]{Nelson2000}
A.~F. {Nelson}.
\newblock {Planet Formation is Unlikely in Equal-Mass Binary Systems with A
  \~{} 50 AU}.
\newblock \emph{\apjl}, 537:\penalty0 L65--L68, July 2000.

\bibitem[{Pfahl}(2005)]{Pfahl2005}
E.~{Pfahl}.
\newblock {Cluster Origin of the Triple Star HD 188753 and Its Planet}.
\newblock \emph{\apjl}, 635:\penalty0 L89--L92, December 2005.
\newblock \doi{10.1086/499162}.

\bibitem[{Pfahl} and {Muterspaugh}(2006)]{PfahlMute2006}
E.~{Pfahl} and M.~{Muterspaugh}.
\newblock {Impact of Stellar Dynamics on the Frequency of Giant Planets in
  Close Binaries}.
\newblock \emph{\apj}, 652:\penalty0 1694--1697, December 2006.
\newblock \doi{10.1086/508446}.

\bibitem[{Pichardo} et~al.(2005){Pichardo}, {Sparke}, and
  {Aguilar}]{Pichardo2005}
B.~{Pichardo}, L.~S. {Sparke}, and L.~A. {Aguilar}.
\newblock {Circumstellar and circumbinary discs in eccentric stellar binaries}.
\newblock \emph{\mnras}, 359:\penalty0 521--530, May 2005.
\newblock \doi{10.1111/j.1365-2966.2005.08905.x}.

\bibitem[{Pilat-Lohinger} and {Dvorak}(2002)]{PL2002}
E.~{Pilat-Lohinger} and R.~{Dvorak}.
\newblock {Stability of S-type Orbits in Binaries}.
\newblock \emph{Celestial Mechanics and Dynamical Astronomy}, 82:\penalty0
  143--153, 2002.

\bibitem[{Pilat-Lohinger} et~al.(2003){Pilat-Lohinger}, {Funk}, and
  {Dvorak}]{PL2003}
E.~{Pilat-Lohinger}, B.~{Funk}, and R.~{Dvorak}.
\newblock {Stability limits in double stars. A study of inclined planetary
  orbits}.
\newblock \emph{\aap}, 400:\penalty0 1085--1094, March 2003.

\bibitem[{Portegies Zwart} and {McMillan}(2005)]{Portegies_Zwart_2005}
S.~F. {Portegies Zwart} and S.~L.~W. {McMillan}.
\newblock {Planets in Triple Star Systems: The Case of HD 188753}.
\newblock \emph{\apjl}, 633:\penalty0 L141--L144, November 2005.
\newblock \doi{10.1086/498302}.

\bibitem[{Queloz} et~al.(2000){Queloz}, {Mayor}, {Weber}, {Bl{\'e}cha},
  {Burnet}, {Confino}, {Naef}, {Pepe}, {Santos}, and {Udry}]{Queloz2000}
D.~{Queloz}, M.~{Mayor}, L.~{Weber}, A.~{Bl{\'e}cha}, M.~{Burnet},
  B.~{Confino}, D.~{Naef}, F.~{Pepe}, N.~{Santos}, and S.~{Udry}.
\newblock {The CORALIE survey for southern extra-solar planets. I. A planet
  orbiting the star Gliese 86}.
\newblock \emph{\aap}, 354:\penalty0 99--102, February 2000.

\bibitem[{Rabl} and {Dvorak}(1988)]{Rabl1988}
G.~{Rabl} and R.~{Dvorak}.
\newblock {Satellite-type planetary orbits in double stars - A numerical
  approach}.
\newblock \emph{\aap}, 191:\penalty0 385--391, February 1988.

\bibitem[{Shao} and {Colavita}(1992)]{shao92}
M.~{Shao} and M.~M. {Colavita}.
\newblock {Potential of long-baseline infrared interferometry for narrow-angle
  astrometry}.
\newblock \emph{\aap}, 262:\penalty0 353--358, August 1992.

\bibitem[{Shao} and {Staelin}(1980)]{ss80}
M.~{Shao} and D.~H. {Staelin}.
\newblock {First fringe measurements with a phase-tracking stellar
  interferometer}.
\newblock \emph{\ao}, 19:\penalty0 1519--1522, May 1980.

\bibitem[{Th{\'e}bault} et~al.(2004){Th{\'e}bault}, {Marzari}, {Scholl},
  {Turrini}, and {Barbieri}]{thebault2004}
P.~{Th{\'e}bault}, F.~{Marzari}, H.~{Scholl}, D.~{Turrini}, and M.~{Barbieri}.
\newblock {Planetary formation in the {$\gamma$} Cephei system}.
\newblock \emph{\aap}, 427:\penalty0 1097--1104, December 2004.
\newblock \doi{10.1051/0004-6361:20040514}.

\bibitem[{Th{\'e}bault} et~al.(2006){Th{\'e}bault}, {Marzari}, and
  {Scholl}]{Thebault2006}
P.~{Th{\'e}bault}, F.~{Marzari}, and H.~{Scholl}.
\newblock {Relative velocities among accreting planetesimals in binary systems:
  The circumprimary case}.
\newblock \emph{Icarus}, 183:\penalty0 193--206, July 2006.
\newblock \doi{10.1016/j.icarus.2006.01.022}.

\bibitem[{Toyota} et~al.(2005){Toyota}, {Itoh}, {Matsuyama}, {Urakawa},
  {Kimura}, {Oasa}, {Mukai}, and {Sato}]{Toyota2005}
E.~{Toyota}, Y.~{Itoh}, H.~{Matsuyama}, S.~{Urakawa}, S.~{Kimura}, Y.~{Oasa},
  T.~{Mukai}, and B.~{Sato}.
\newblock {Search for Extrasolar Planets in Binary Systems}.
\newblock pages 8247--+, 2005.

\bibitem[{Udry} et~al.(2004){Udry}, {Eggenberger}, {Mayor}, {Mazeh}, and
  {Zucker}]{Udry2004}
S.~{Udry}, A.~{Eggenberger}, M.~{Mayor}, T.~{Mazeh}, and S.~{Zucker}.
\newblock {Planets in multiple-star systems:properties and detections}.
\newblock pages 207--214, August 2004.

\bibitem[{Watson} and {Dhillon}(2004)]{Starspots}
C.~A. {Watson} and V.~S. {Dhillon}.
\newblock {The effect of star-spots on eclipse timings of binary stars}.
\newblock \emph{\mnras}, 351:\penalty0 110--116, June 2004.

\bibitem[{Woltjer}(1922)]{woltjer1922}
J.~{Woltjer}.
\newblock {On a special case of orbit determination in the theory of eclipsing
  variables}.
\newblock \emph{\bain}, 1:\penalty0 93--+, June 1922.

\bibitem[{Young}(1967)]{Young1967}
A.~T. {Young}.
\newblock {Photometric error analysis. VI. Confirmation of Reiger's theory of
  scintillation}.
\newblock \emph{\aj}, 72:\penalty0 747--+, August 1967.

\bibitem[{Zucker} and {Mazeh}(1994)]{Zucker1994}
S.~{Zucker} and T.~{Mazeh}.
\newblock {Study of spectroscopic binaries with TODCOR. 1: A new
  two-dimensional correlation algorithm to derive the radial velocities of the
  two components}.
\newblock \emph{\apj}, 420:\penalty0 806--810, January 1994.
\newblock \doi{10.1086/173605}.

\bibitem[{Zucker} et~al.(2004){Zucker}, {Mazeh}, {Santos}, {Udry}, and
  {Mayor}]{Zuc2004}
S.~{Zucker}, T.~{Mazeh}, N.~C. {Santos}, S.~{Udry}, and M.~{Mayor}.
\newblock {Multi-order TODCOR: Application to observations taken with the
  CORALIE echelle spectrograph. II. A planet in the system HD 41004}.
\newblock \emph{\aap}, 426:\penalty0 695--698, November 2004.
\newblock \doi{10.1051/0004-6361:20040384}.

\end{thebibliography}
\bibliographystyle{plainnat}
\end{document}